\documentclass[aps,prb,twocolumn,groupedaddress,showkeys,showpacs]{revtex4}

\usepackage{multirow,amssymb,amsbsy,amsmath}
\usepackage{epsfig}

\newcommand {\nifour} {Ni$_{4}$}
\newcommand {\nimo} {\{Ni$_{4}$Mo$_{12}$\}}
\newcommand {\nitwo} {Ni$^{\text{II}}$}
\newcommand {\nifourfull} {[Mo$^{\text{V}}_{12}$O$_{30}$($\mu_2$-OH)$_{10}$H$_2$\{Ni$^{\text{II}}$(H$_2$O)$_3$\}$_4$]}
\newcommand{\op}[1]{%
    \fontdimen12\textfont3=2pt\fontdimen12\scriptfont3=1.4pt%
    \!\null\mathop{\vphantom{#1}\smash{#1}}\limits_{\sim}\null\!}
\newcommand{\xref}[1]{\protect\ref{#1}}
\newcommand{\figref}[1]{Fig.~\protect\ref{#1}}
\newcommand{\fmref}[1]{(\protect\ref{#1})}

\newcommand{\dint}{\text{d}}

\begin{document}

\title{Observation of field-dependent magnetic parameters
  in the magnetic molecule \nimo}

\author{J\"urgen Schnack}
\email{jschnack@uos.de}
\author{Mirko Br\"uger}
\affiliation{Universit\"at Osnabr\"uck, Fachbereich Physik,
D-49069 Osnabr\"uck, Germany}

\author{Marshall Luban}
\author{Paul K\"ogerler}
\author{Emilia Morosan}
\author{Ronald Fuchs}
\affiliation{Ames Laboratory \& Department of Physics and Astronomy,
Iowa State University, Ames, Iowa 50011, USA}

\author{Robert Modler}
\affiliation{Johann Modler GmbH, Postfach 100462, D-63741
  Aschaffenburg, Germany}

\author{Hiroyuki Nojiri}
\affiliation{Institute for Materials Research, Tohoku
  University, Katahira 2-1-1, Sendai 980-8577, Japan}

\author{Ram C. Rai}
\author{Jinbo Cao}
\author{Janice L. Musfeldt}
\affiliation{Department of Chemistry, University of Tennessee,
  Knoxville, TN 37996, USA}

\author{Xing Wei}
\affiliation{National High Magnetic Field Laboratory, Florida State University,
Tallahassee, FL, 32310}

\date{\today}

\begin{abstract}
  We investigate the bulk magnetic, electron paramagnetic
  resonance, and magneto-optical properties of \nimo, a magnetic
  molecule with antiferromagnetically coupled tetrahedral
  \nitwo\ in a diamagnetic molybdenum matrix. The
  low-temperature magnetization exhibits steps at irregular
  field intervals, a result that cannot be explained using a
  Heisenberg model even if it is augmented by magnetic
  anisotropy and biquadratic terms.  Allowing the exchange and
  anisotropy parameters to depend on the magnetic field provides
  the best fit to our data, suggesting that the molecular
  structure (and thus the interactions between spins) may be
  changing with applied magnetic field.
\end{abstract}

\maketitle

\section{Introduction}
\label{sec-1}

Despite the enormous progress that has been made in
understanding magnetic molecules over the past
decade,\cite{SGC:Nat93,GCR:S94,Gat:AM94,Cor:NATO96,CGS:JMMM99,WSK:IO99,MKD:CCR01,MLS:CPC01,WAH:Nature02}
it is still a challenge to deduce the underlying microscopic
spin Hamiltonian. Mn$_{12}$-acetate is a good example of this
problem.\cite{Lis:ACB80,SGC:Nat93,TLB:Nature96,Cor:NATO96,JLB:PRL00,FWK:PRB01}
Although known for almost twenty years, only extensive
investigation elucidated the model
parameters.\cite{RJS:PRB02,CSO:PRB04} Small magnetic molecules,
with their simpler chemical and magnetic structure, can
therefore provide an important opportunity to understand the
dependence of magnetic observables on model
parameters.\cite{WHM:PRB98,KWM:PRB03,LBB:PRB03}

In this work we report our joint experimental and theoretical
efforts to understand the behavior of \nifourfull, henceforth
abbreviated as \nimo, a magnetic molecule which is comprised of
\nitwo\ centers positioned at the nucleophilic sites of an
$\epsilon$-Keggin cluster forming an almost ideal
tetrahedron.\cite{MBK:IC00} In contrast to several other nickel
compounds which exhibit
ferromagnetic\cite{ABB:CEJ02,KWM:PRB03,YWH:P03,MHR:EJIC04,KCW:DT04}
or mixed coupling,\cite{DLM:IC05} the Ni centers of this molecule
are antiferromagnetically coupled, as is also the case for certain
Ni-$2\times 2$-grid molecules.\cite{WHM:PRB98} Because the structure
of \nimo\ is almost perfectly tetrahedral one might anticipate that
the magnetic energy levels are reasonably well described by an
isotropic Heisenberg model with a single antiferromagnetic exchange
parameter.\cite{MBK:IC00} Such a Hamiltonian can be expressed in
terms of the total spin:
\begin{eqnarray}
\label{E-1-1}
\op{H}
&=&
-2 J
\sum_{u<v}\;
\op{\vec{s}}(u) \cdot \op{\vec{s}}(v)
+
g \mu_B \vec{B}\cdot \sum_{u}\;\op{\vec{s}}(u)
\\
&=&
-J
\left[
\op{\vec{S}}^2-4s(s+1)
\right]
+
g \mu_B B \;\op{S}_z
\nonumber
\ ,
\end{eqnarray}
where $\op{\vec{s}}(u)$ is a single-spin operator at site $u$
and $\op{\vec{S}}$ is the total spin operator. The spin quantum
number of each \nitwo\ ion in an octahedral ligand field is
$s=1$. For antiferromagnetic coupling ($J<0$) the resulting
low-temperature magnetization curve $\mathcal{M}(B)$ that
follows from \fmref{E-1-1} displays four steps before reaching
saturation. These steps occur at the magnetic fields
\begin{eqnarray}
\label{E-1-2}
B_{S\rightarrow(S+1)}
&=&
-
\frac{2 J}{g \mu_B} (S+1)
\ ,
\end{eqnarray}
for $S=0, 1, 2, 3$, where the lowest Zeeman-split levels of
adjacent multiplets cross. In particular, from \fmref{E-1-2} it
follows that the level crossing fields are uniformly
spaced.\cite{ACC:ICA00,ScL:PRB01} In stark contrast to the
expectation of uniformly spaced crossing fields, the
experimental magnetization ${\mathcal M}(B)$ curve of \nimo\
features non-equidistant steps at 4.5, 8.9, 20.1, and 32~T. Even
assuming an anisotropic Hamiltonian\cite{Brueger03} with two
exchange couplings and biquadratic terms as done for other
\nitwo-compounds,\cite{WHM:PRB98,KWM:PRB03} we were unable to
account for the specific sequence of steps in the
low-temperature magnetization of \nimo.  In order to provide a
comprehensive picture of the unusual high-field magnetic
behavior of \nimo, we have been led to invoke field-dependent
exchange and anisotropy parameters.  We argue that this
dependence emanates from changes in molecular structure (and
thus the overlap of those atomic orbitals that determine the
exchange interactions as well as the coordination geometries
which affect the electronic single-ion properties) with applied
magnetic field. The magneto-optical response of \nimo\ supports
a small change in the \nitwo\ coordination environment and the
associated electronic single-ion properties.

\section{Chemical and Crystal Structure}
\label{sec-6}

\begin{figure}[!ht]
\begin{center}
\epsfig{file=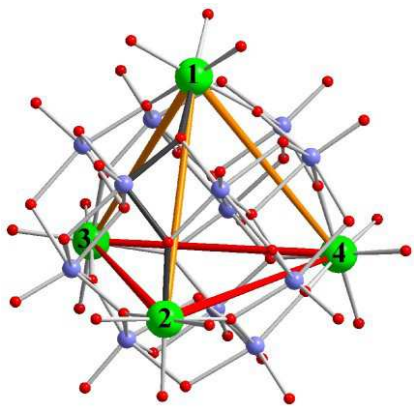,width=55mm}
\vspace*{1mm}
\caption[]{(color online) Ball-and-stick representation of the
  \nimo\ molecule (Ni: numbered big spheres; Mo: medium size
  spheres; O: small spheres; H: not shown) emphasizing a
  slightly stretched Ni$^{\text{II}}_4$ pyramid with a
  near-equilateral triangle base (Ni ions labeled 2, 3, and 4)
  and an elevated apex (Ni ion labeled 1).  Ni-Ni distances:
  $d_{12}=6.700(5)$~\AA, $d_{13}=d_{14}=6.689(1)$~\AA,
  $d_{23}=d_{24}=6.616(1)$~\AA,
  $d_{34}=6.604(1)$~\AA.\cite{MBK:IC00}}
\label{F-1}
\end{center}
\end{figure}

\begin{figure}[!ht]
\begin{center}
\epsfig{file=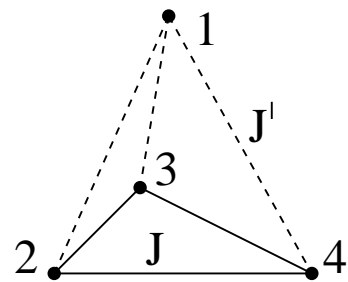,width=45mm}
\vspace*{1mm}
\caption[]{Simplified structure of \nifour: the superexchange
  interactions $J'$ and $J$ are represented by dashed and solid
  lines, respectively.}
\label{F-2}
\end{center}
\end{figure}

The neutral \nimo\ cluster is isolated in the form of crystals
of
[Mo$^{\text{V}}_{12}$O$_{30}$($\mu_2$-OH)$_{10}$H$_2$\{Ni$^{\text{II}}$(H$_2$O)$_3$\}$_4$]$\cdot$14~H$_2$O
and is based on the diamagnetic, highly-charged
$\epsilon$-Keggin anion
[Mo$^{\text{V}}_{12}$O$_{38}$($\mu_3$-OH)$_{2}$]$^{18-}$, built
up from four edge-sharing \{Mo$_3$\} groups (each consisting of
three edge-sharing MoO$_6$ octahedra).  Within the
$\epsilon$-Keggin framework, the Mo positions form six
Mo$^{\text{V}}_2$ groups with short Mo-Mo single bonds.  The
$\epsilon$-Keggin structure is formally derived from the common
$\alpha$-Keggin isomer by rotating all four {Mo$_3$} groups by
60$^\circ$, preserving the T$_d$ symmetry. In \nimo, four
[\nitwo(H$_2$O)$_3$]$^{2+}$ groups are coordinated each to three
(unprotonated) $\mu_2$-oxo centers that interlink the Mo
positions of the Mo$^{\text{V}}_2$ groups (\figref{F-1}). This
results in a octahedral O$_3$\nitwo(H$_2$O)$_3$ coordination
environment with all-trans-positioned oxo and water ligands and
nearly identical Ni-O distances (Ni-($\mu_2$-O):~2.05~\AA,
Ni-OH$_2$:~2.06~\AA).  This capping of the {Mo$_3$} groups of
the $\epsilon$-Keggin fragment by four \nitwo\ positions
produces a near-regular \nifour\ tetrahedron. Contrary to many
other tetrahedral \nifour\ structures in which Ni pairs are
connected by mononuclear bridging centers, in \nimo\ each Ni
pair is interconnected via one -O(-Mo-)$_2$O- bridging motif
serving as a superexchange pathway (see \figref{F-1} where one
Ni-O-Mo-O-Ni pathway is highlighted by dark bonds).  The
geometry of each of these pathways is therefore characterized by
four bond lengths, three bond angles, and two dihedral angles,
as opposed to \nifour-type structures comprising mononuclear
linker groups (two bond lengths, one bond angle).  Importantly
however, the molecular geometry of \nimo\ slightly deviates from
the $T_d$-symmetric ideal, resulting in a slightly stretched
\nifour\ pyramid with elongated Ni-Ni distances between the Ni
positions of a basal Ni$_3$ plane and one apex.  While the
crystallographic symmetry operations result in the molecular
point group $C_s$, the actual geometry is virtually of $C_{3v}$
symmetry, and the geometric parameters for all
Ni-O(-Mo-)$_2$O-Ni pathways correspondingly fall into two sets.
Within these two sets, the individual bond lengths and angles
display minimal deviations (typically $< 0.8$~\%) from the
respective averages: Intra-basal Ni-Ni contacts are
characterized by $\langle$Ni-O$\rangle$ = 2.05\AA,
$\langle$Mo-O$\rangle$ = 1.95~\AA, $\langle$Ni-O-Mo$\rangle$ =
135.4$^\circ$, and $\langle$Mo-O-Mo$\rangle$ = 89.6$^\circ$. For
Ni-Ni contacts between the apex to the base positions average
values of $\langle$Ni-O$\rangle$ = 2.03~\AA,
$\langle$Mo-O$\rangle$ = 1.95~\AA, $\langle$Ni-O-Mo$\rangle$ =
137.9$^\circ$, and$\langle$<Mo-O-Mo$\rangle$ = 94.6$^\circ$ are
found.  As the geometric parameters do not vary significantly
within the intra-basal and within the apex-basal Ni-Ni contacts, 
we do not take into account slight deviations from the idealized
$C_{3v}$ symmetry, and use only exchange constants $J$ and $J'$
in this paper (see \figref{F-2}).

It should be added that magnetic
exchange through polyoxomolybdates frameworks, especially if
more than one possible pathway exists for each contact, has been
found to be fairly insensitive to the separation distance of the
pair of spin centers. Also, for similar systems based on
mononuclear linker groups the bond angles have a stronger effect
on the exchange energies than the contact
distance.\cite{HSH:IC95,WHM:PRB98,EFK:P99,CMP:CC01,ABB:CEJ02,KWM:PRB03,YWH:P03,MHR:EJIC04}
Due to the presence of crystal water molecules in the
solid-state structure of \nimo$\cdot$14~H$_2$O which space the
cluster entities apart, the closest intermolecular Ni$\cdots$Ni
distance in the solid state exceeds 7.15~\AA, rendering
inter-molecular (dipole-dipole) magnetic exchange insignificant.

\section{Experimental results}
\label{sec-2}

\subsection{Magnetic properties of \nimo}
\label{sec-2-1}

\begin{figure}[!ht]
\begin{center}
\epsfig{file=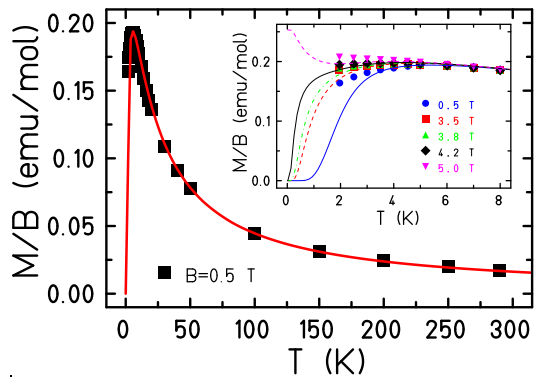,width=70mm}
\vspace*{1mm}
\caption[]{(color online) Low-field susceptibility: Experimental
  data are given by black squares whereas the solid curve shows
  the result assuming a simple Heisenberg Hamiltonian
  \fmref{E-1-1} with $J/k_B=-3.4$~K and an isotropic
  spectroscopic splitting factor $g=2.25$. The inset shows a
  close-up view of the low-temperature region for different
  magnetic fields.}
\label{F-3}
\end{center}
\end{figure}

Figure~\xref{F-3} displays the magnetic susceptibility
$\mathcal{M}(B)$ of \nimo\ as a function of temperature.  These
data were collected on a SQUID magnetometer (Quantum Design
MPMS-5) at various magnetic fields in a temperature range of $2
- 290$~K. Using a simple Heisenberg Hamiltonian \fmref{E-1-1}
one obtains $J/k_B=-3.4$~K and an isotropic $g$-factor of
$g=2.25$, see also Ref.~\onlinecite{MBK:IC00}.  The model fit
given by the solid curve in \figref{F-3} is acceptable, although
the low-temperature behavior is not well reproduced, as can be
seen in the inset of \figref{F-3}, where the magnetic
susceptibility for various field values and low temperatures is
displayed. We suggest that the deviation is not only due to the
presence of impurities but also due to anisotropic as well as
biquadratic terms in the Hamiltonian which are known to be
needed to model the low-temperature behavior of
\nitwo-compounds.\cite{WHM:PRB98,KWM:PRB03}

\begin{figure}[!ht]
\begin{center}
\epsfig{file=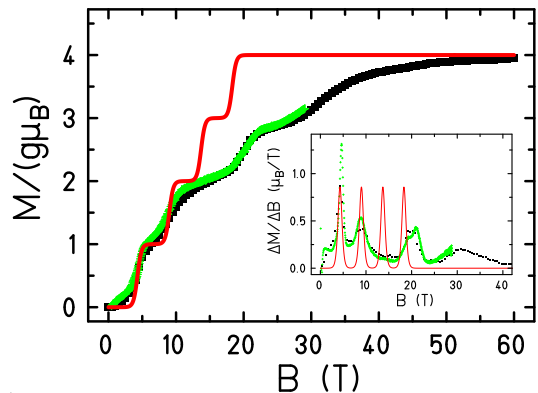,width=70mm}
\vspace*{1mm}
\caption[]{(color online) Magnetization: Experimental data are
  given by squares (dark symbols -- NHMFL, $T=0.44$~K; light
  symbols -- OHMFL, $T=0.40$~K). The theoretical magnetization
  assuming a simple Heisenberg Hamiltonian \fmref{E-1-1} with
  $J/k_B=-3.4$~K and an isotropic spectroscopic splitting factor
  $g=2.25$ is given by a solid curve for $T=0.44$~K. The inset
  shows the experimental differential magnetization $\dint
  {\mathcal M}/\dint B$ as data points as well as the
  theoretical $\dint {\mathcal M}/\dint B$ (solid curve) using
  the same parameters as above.  }
\label{F-4}
\end{center}
\end{figure}

High-field magnetization measurements are a valuable tool to
extract information on the spin Hamiltonian which is not
accessible from magnetization measurements on commercial SQUID
magnetometers.  The high-field magnetization for a powder sample
of \nimo\ has been independently measured in pulsed magnetic
fields at the facility of the National High Magnetic Field
Laboratory (NHMFL) at Los Alamos as well as at the Okayama High
Magnetic Field Laboratory (OHMFL) by using a standard inductive
method (maximum at NHMFL $B=60$~T, whereas the maximum at OHMFL
is $B=40$~T, $dB/dt=10000\dots 15000$~T/s).  The results of
these two measurements are in very close agreement. No
hysteresis is found between up and down sweep runs, indicating
thermal equilibrium behavior of $\mathcal{M}$ vs. $B$.

Figure \xref{F-4} shows the magnetization as a function of
applied external magnetic field at $T=0.44$~K.  Strikingly, four
non-equidistant steps are observed in the magnetization. These
steps are found near 4.5, 8.9, 20.1, and 32~T.  Saturation of
the magnetization is not observed until 60~T. For comparison, we
show the expected response of a simple Heisenberg Hamiltonian
(Eq.~\fmref{E-1-1}, ideal tetrahedron) for the model parameters
extracted above. Note that this model predicts equidistant steps
in the magnetization at 4.5, 9.0, 13.5, and 18.0~T with
saturation above the fourth step. The drastic deviation between
the predictions of Eqn.~\fmref{E-1-1} and the experimental
results cannot be the result of heating via the magnetocaloric
effect. Although observed in other compounds, such an effect
would only smear out the steps but not shift the step positions.
In addition sample heating or cooling in a varying field is
often accompanied by hysteresis, which is absent in this
measurement.\footnote{The interested reader is referred to the
  extensive literature on ``butterfly hysteresis loops'' in
  molecular magnets, e.g.
  Refs.~\onlinecite{CWM:PRL00,WKS:PRL02,WKS:PRL03}.}  The
observed step positions thus constitute a challenge not only to
the simple Heisenberg model given by \eqref{E-1-1}, but also to
more elaborate models \eqref{E-3-1-1} incorporating anisotropy
terms and biquadratic exchange. We suggest that these models
should be extended to include field-dependent parameters in
order to account for our experimental results.  This will be
discussed in Sec.~\xref{sec-4}.

\subsection{EPR response of \nimo}
\label{sec-2-3}

\begin{figure}[!ht]
\begin{center}
\epsfig{file=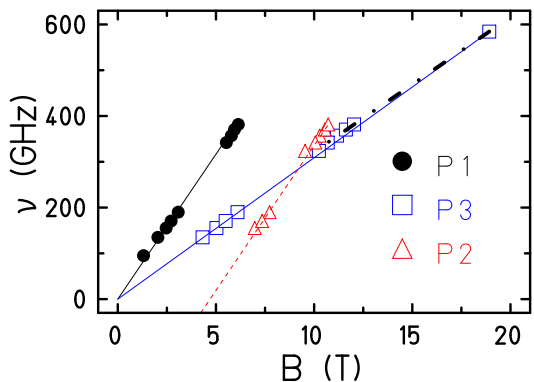,width=70mm}
\vspace*{1mm}
\caption[]{(color online) The figure shows the dependence of the
  observed EPR resonance frequencies P1 (circles), P2
  (triangles), and P3 (squares) on the magnetic field. The lines
  provide linear fits to the data. The data were taken at
  $T=4.2$~K.}
\label{F-5}
\end{center}
\end{figure}

\begin{figure}[!ht]
\begin{center}
\epsfig{file=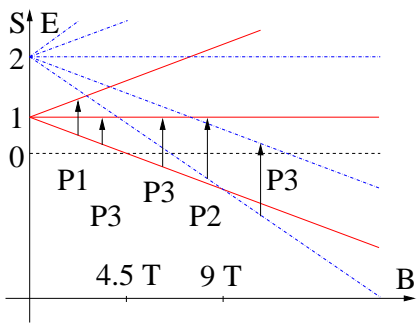,width=55mm}
\vspace*{1mm}
\caption[]{(color online) The figure shows schematically the
  Zeeman level splittings in a pure Heisenberg model together
  with the assignments of allowed and forbidden transitions.  }
\label{F-6}
\end{center}
\end{figure}

Figure \xref{F-5} displays the results of our Electron
Paramagnetic Resonance (EPR) measurements. The transmission of a
powder sample was determined as a function of applied magnetic
field in a frequency range from 95~GHz to 381.5~GHz. The figure
shows the dependence of the observed EPR resonance frequencies
P1 (circles), P2 (triangles), and P3 (squares) on magnetic
field. One immediately notices that two different slopes can be
assigned to the data, one corresponding to $\Delta M = 1$ and
another one corresponding to forbidden transitions with $\Delta
M = 2$. These dependencies can be qualitatively -- and to some
extent quantitatively -- explained by looking at the Zeeman
level scheme of the simple Heisenberg model \fmref{E-1-1} as it
is schematically depicted in \figref{F-6}.

The strongest transition which we observe in the spectra is the
allowed transition P3 with $\Delta M = 1$. Since the applied
temperature is of the order of the coupling, this transition is
actually a sum of several transitions. At low-field values it is
dominated by the transition between ($S=1, M=-1$) and ($S=1,
M=0$), whereas at higher fields the dominant contribution stems
from the transition between ($S=2, M=-2$) and ($S=2, M=-1$).
The dependence of the resonance frequency of P3 on the applied
field suggests that the zero-field splitting in the triplet is
small.

P1 is a low-field transition which connects ($S=1, M=-1$) and
($S=1, M=+1$) and thus should be forbidden. In the spectra its
strength is much weaker than that of P3. We believe that this
transition appears due to mixing of $\op{S}_z$ eigenstates that would
arise from anisotropic contributions to the Hamiltonian. The
line which is plotted through the data points suggests that the
zero-field splitting between ($S=1, M=-1$) and ($S=1, M=+1$)
is probably small, although, the experimental data points --
which extend only down to 1.3~T -- would also allow a somewhat
bigger splitting, especially since the lowest-lying P1 data
points appear to deviate from a straight line.

P2 is another rather weak forbidden transition which shares the
slope with P1. We believe that this transition connects ($S=2,
M=-2$) and ($S=1, M=0$). This transition is not observable below
about 7~T due to the fact that this transition occurs only when
the mixing of $\op{S}_z$ eigenstates is sufficiently strong
which is the case around the level crossing at 9~T.  The
dependence of this transition on temperature, i.e. on the
thermal occupation of the level with ($S=2, M=-2$) is small. The
extrapolation of the field dependence allows one to deduce an
approximate isotropic Heisenberg coupling from the zero-field
energy separation of the triplet and the pentuplet. From
Eq.~\fmref{E-1-1} one deduces that $E(S=2)-E(S=1)=-4J$, thus one
obtains $J\approx -3.4$~K, which is in very good agreement with
the exchange constants deduced from our susceptibility
measurements. The spectroscopic splitting factor can be
determined from the slopes of P3 and P1 in \figref{F-5} to be
$g=2.23\pm 0.03$, which is also in good agreement with the value
deduced from the high-temperature behavior of the
susceptibility. It is noteworthy that the high field data of
transition P3 have a smaller slope which, if fitted alone
(dashed-dotted line in \figref{F-5}), suggests $g\approx 2.11\pm
0.03$. However, we want to point out that these considerations
are done on the basis of the simple Heisenberg model
\fmref{E-1-1}, that does not take into account that a realistic
Hamiltonian has to contain anisotropic terms.  A more detailed
explanation is given at the end of sec.~\xref{sec-4}.

Summarizing this part, we find that the zero-field splitting is
small. We explain this observation by the fact that the local
principal axes of the Ni-ions point in different directions
(radially outwards), and thus the average global anisotropy is
small. Our use of a powder sample led to additional averaging.

\subsection{Optical properties of \nimo}
\label{sec-2-4}

\begin{figure}[!ht]
\includegraphics[width=80mm]{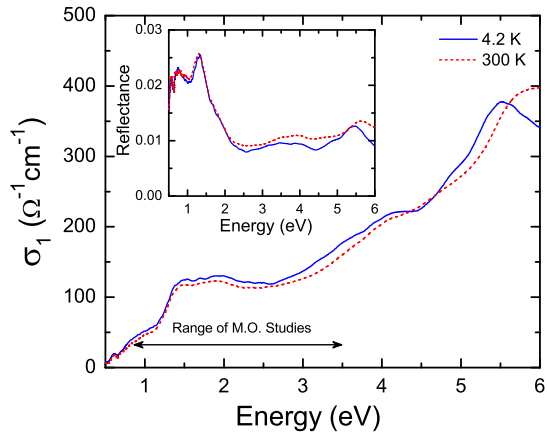}
\vskip -0.2cm \caption{(color online) Optical conductivity of a
  \nimo\ pellet at 4.2~K (solid curve) and 300~K (dashed curve),
  calculated from reflectance measurements (inset) by
  Kramers-Kronig analysis. The energy range of our
  magneto-optical work is indicated by the arrow.}
\label{F-7}
\end{figure}

Figure \ref{F-7} displays the optical conductivity of \nimo.
These experiments were carried out on a pressed powder sample
using a Lambda-900 grating spectrometer equipped with
reflectance stage and cryostat.\cite{CWM:PRB04} \nimo\ is a
semiconductor with an optical gap of $\sim$0.6 eV. Based upon
comparisons with chemically similar nicklates as well as
existing electronic calculations,\cite{PYH:JAP05,PBS:PT05} we
assign the excitations centered at 1.5 eV as on-site nickel $d$
to $d$ transitions, likely activated by modest hybridization
with the coordinating oxygens.  These excitations take place in
both majority and minority channels according to recent DFT
calculations.\cite{PBS:PT05} The features above 3 eV are
assigned as O $p$ to Ni $d$ charge transfer excitations. The
energy range of our magneto-optical investigation is also shown
in Fig. \ref{F-7}, providing a preview of the physical origin
of the field-induced spectroscopic changes, discussed below.

\begin{figure}[!ht]
\includegraphics[width=80mm]{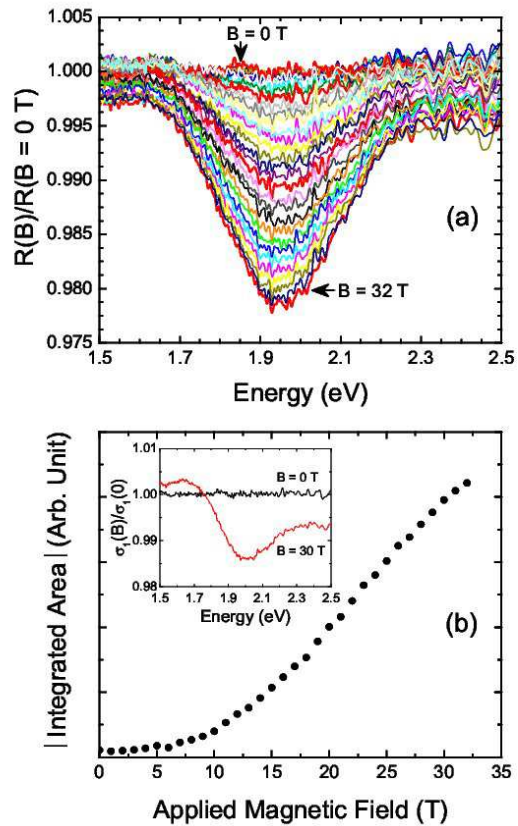}
\vskip -0.2cm
\caption{(color online) (a) The normalized magneto-optical
  response, R(B)/R(B = 0 T), of \nimo\ pressed powder in an
  applied magnetic field from 0 to 32 T at 4.2 K. 1 T steps are
  shown. No hysteresis is observed.  (b) Absolute value of
  the integrated area of the magneto-optical reflectance ratio
  feature as a function of applied magnetic field (solid
  symbol).  Results from the integrated optical conductivity
  data are similar.  The inset shows the change in the optical
  conductivity of \nimo\ with magnetic field, $\sigma_1$($B$ =
  30 T)/$\sigma_1$($B$ = 0 T).  The $B$ = 0 T curve ratios data
  taken before and after the field sweep, giving an indication
  of overall spectral reproducibility.}
\label{F-8}
\end{figure}

Figure \ref{F-8}(a) displays the magneto-optical response of
\nimo\ as a function of magnetic field from 0 to 32~T at 4.2~K.
These experiments were carried out on the same pressed powder
sample using a grating spectrometer (0.8 - 3.5 eV) equipped with
InGaAs and CCD detectors and a 33~T steady field resistive
magnet at the NHMFL in Tallahassee, FL.\cite{CWM:PRB04} The
reflectance ratio, $R(B)/R(B = 0)$, is a normalized response and
highlights changes in the optical properties with applied
magnetic field.  With increasing field, the reflectance of
\nimo\ decreases by $\sim$2\%. It is notable that this effect
occurs in the visible spectral range, hence the name
``magnetochromism". Based upon the aforementioned peak
assignments in the optical conductivity spectrum in \figref{F-7}
and the $\sigma_1$($B$)/$\sigma_1$($B$ = 0 T) ratio data in the
inset of \figref{F-8}(b), we attribute the observed
magnetochromic effect to a field-induced modification of the Ni
$d$ to $d$ on-site excitation. Distortion of the
pseudo-octahedral \nitwo\ crystal field environment is a
plausible driver. Note that this is a local, molecular-level
distortion rather than a bulk effect.  No field-induced changes
were observed on the leading edge of the O $p$ to Ni $d$ charge
transfer bands above 3~eV.

We quantify the magneto-optical effect in \nimo\ by plotting the
absolute value of the integrated intensity of the reflectance
ratio as a function of applied magnetic field
(\figref{F-8}(b)).\footnote{Similar results are obtained using
  the optical conductivity ratio, $\sigma_1(B)/\sigma_1(0)$.}
The magneto-optical effect is small at low fields, becomes
appreciable above 10~T, and continues to grow for $B\ge 30$~T.
There is no evidence of saturation to 33 T. The overall rising
magnetochromic response can be fit with several different
functions including a cubic polynomial in $|B|$ and a simple
exponential, suggesting a likely functional form for the
field-dependence of the magnetic parameters in \nimo, described
below.  We propose that the applied magnetic field interacts
with the spin centers, deforming the local structure around the
\nitwo\ sites.  This process modifies the crystal field
environment, the result of which is a field-induced change in
electronic structure (\figref{F-8}~(a)).  Thus, these
spectroscopic results support the picture of field-dependent
exchange and anisotropy terms in the spin Hamiltonian of \nimo\
that derive from magnetoelastic (and consequent spin-orbit)
interactions.  Magneto-optical effects due to field-induced
changes in local structure have also been observed in other
materials including (CPA)$_2$CuBr$_2$ and
K$_2$V$_3$O$_8$.\cite{WCM:PRB,rai05}

\section{Theoretical Models}
\label{sec-3}

In order to understand the magnetic properties of \nimo\ we
adopt the following general Hamiltonian that has been used for
other \nitwo-compounds,\cite{WHM:PRB98,KWM:PRB03}
\begin{eqnarray}
\label{E-3-1-1}
\op{H}
&=&
\op{H}_{\text{H}}
+
\op{H}_{\text{ani}}
+
\op{H}_{\text{biq}}
+
\op{H}_{\text{Z}}
\ ,\ \text{where}
\\
\label{E-3-1-1-A}
\op{H}_{\text{H}}
&=&
-
2
\sum_{u<v}\;
J_{uv}
\op{\vec{s}}(u) \cdot \op{\vec{s}}(v)
\\
\label{E-3-1-1-B}
\op{H}_{\text{ani}}
&=&
D
\left[
\sum_{u}\;
(\vec{e}_r(u) \cdot \op{\vec{s}}(u))^2
-\frac{8}{3}
\right]
\\
\label{E-3-1-1-C}
\op{H}_{\text{biq}}
&=&
-
2
\sum_{u<v}\;
j_{uv}
\left(\op{\vec{s}}(u) \cdot \op{\vec{s}}(v)\right)^2
\\
\label{E-3-1-1-D}
\op{H}_{\text{Z}}
&=&
g\, \mu_B\, \vec{B} \cdot
\sum_{u}\;
\op{\vec{s}}(u)
\ .
\end{eqnarray}
Here $\op{H}_{\text{H}}$ denotes the Heisenberg Hamiltonian and
$\op{H}_{\text{ani}}$ describes the single-site anisotropic
(ligand field) contribution, which is compatible with the
approximate tetrahedral symmetry of \nimo.  The unit vector
$\vec{e}_r(u)$, which points radially outwards, serves as a
local anisotropy axis for site $u$. The term $8/3$ is convenient
in order to render the Hamiltonian traceless.
$\op{H}_{\text{biq}}$ represents biquadratic terms, which are
the next higher order compared to the Heisenberg
Hamiltonian,\cite{BeG90} and $\op{H}_{\text{Z}}$ is the Zeeman
term. We employ a single spectroscopic splitting factor $g$
since the $g$-tensor anisotropy was found to be very small for
the present system; a similar result has been found for \nitwo\
squares.\cite{WHM:PRB98} We also assume that a possible
anisotropic exchange between the \nitwo\ centers is small
because the orbital contribution to the ground state is
small.\cite{BBC:CPC03}

In the following sections, we simplify \mbox{$J_{uv} = J$} and
\mbox{$j_{uv} = j$}, or, if we use two different constants,
\mbox{$J' = J_{12} =J_{13} = J_{14}$}, \mbox{$J= J_{23} =J_{24}
  = J_{34} $} and \mbox{$j' = j_{12} =j_{13} = j_{14}$} ,
\mbox{$j= j_{23} =j_{24} = j_{34} $}, as illustrated in
\figref{F-2}. We also account for impurity effects (additional
paramagnetic \nitwo\ ions) and their batch-to-batch variation by
adding a paramagnetic term to the Hamiltonian \fmref{E-3-1-1}.

We have made numerous attempts to model the experimental
magnetization curve of \nimo\ (Fig. \ref{F-4}) with
field-independent values of $J$ and $D$, as described
previously.  Despite these efforts, an explanation for the
non-uniform spacing of the crossing fields has not been
forthcoming.  Assuming a Hamiltonian with two exchange couplings
and anisotropic as well as biquadratic terms as given in
\eqref{E-3-1-1} did not result in a satisfactory description of
all magnetic observables on a common footing.\cite{Brueger03}
Therefore, we extended our model to allow $J$ and $D$ to vary
with applied magnetic field. The possible field-induced
distortion of the crystal field environment around the Ni
centers motivates this ansatz.

\section{Field-dependent model parameters}
\label{sec-4}
\subsection{Low-field properties of \nimo}

It is not \emph{a priori} clear how the model parameters should
depend on the magnetic field since this dependence is indirect.
The exchange couplings, the biquadratic contributions as well as
the anisotropy result from the electronic structure, and are
(complicated) functions of orbital overlaps, lattice stiffness,
and spin-orbit coupling. If the electronic structure is
noticeably altered by an applied field it is probable that the
parameters entering Hamiltonian \fmref{E-3-1-1} are also
changed, but very likely in a highly non-linear manner. To the
best of our knowledge, so far such dependencies were only
observed in certain Mn grid molecules,\cite{WZT:PRL02} where the
anisotropy changes sign at a level crossing.  It appears
plausible to us that besides continuous variations, the
parameters can also change abruptly since the molecular
structure may relax into new ground states at certain field
values. Therefore, one can anticipate that the field-dependence
of the model parameters might only be piecewise analytic between
the abrupt changes. In this section we therefore investigate
simple piecewise parameterizations of the Hamiltonian. We also
neglect a possible anisotropic field dependence of the
parameters, i.e. the exchange parameters of different bonds
would be modified differently, and therefore the effect should
depend on the relative orientation of the molecule with respect
to the field.

\begin{figure}[!ht]
\begin{center}
\epsfig{file=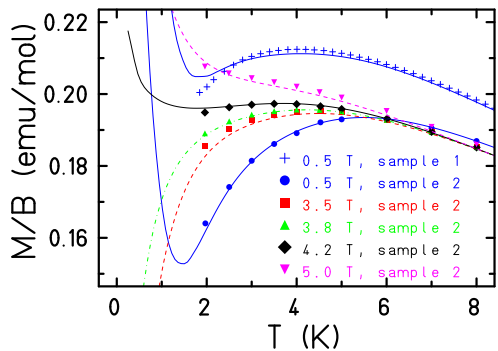,width=65mm}
\vspace*{1mm}
\caption[]{(color online) Low-field susceptibility: The measured
  data are given by symbols.  The fits for $B=0.5$~T are
  obtained using $J/k_B=-3.2$~K, $J'/k_B=-3.1$~K, $j/k_B=1.6$~K,
  $j'/k_B=0$, $D/k_B=-1.0$~K. The data for field values of
  $B=3.5$~T up to $5$~T are approximated using $J/K=-3.2$~K,
  $J'/k_B=-3.2$~K, $j/k_B=1.5$~K, $j'/k_B=0$, and
  $D/k_B=-3.2$~K. $g=2.195$ in all cases. The estimated impurity
  concentrations are 0.216 and 0.146 individual \nitwo\ ions per
  \nimo\ molecule for sample 1 and 2, respectively.  The
  simulated data are averaged over 100 orientations.}
\label{F-9}
\end{center}
\end{figure}

It turns out that the low-field susceptibility versus
temperature data, shown in \figref{F-9}, can be modeled by two
sets of parameters. The magnetization of two different samples
of \nimo\ at $B=0.5$~T can be understood assuming
$J/k_B=-3.2$~K, $J'/k_B=-3.2$~K, $j/k_B=1.6$~K, $j'=0$,
$D/k_B=-1.0$~K together with the separate term to account for
the paramagnetic impurities (free \nitwo\ ions) contained in the
samples (parameterization \textbf{1}). The parameters of the
Hamiltonian compare nicely to those of $2\times 2$--\nitwo\ grid
molecules.\cite{WHM:PRB98} The sign of the anisotropy as well as
that of the biquadratic term signal the same behavior, only the
absolute value of the biquadratic term is somewhat larger.  The
resulting zero-field splitting of the triplet is only 0.15~K, in
very good agreement with our EPR results. The fact that the
biquadratic term is one order of magnitude larger than usually
observed ($j\propto J/100$)\cite{HuO:PRL64,BeG90,WHM:PRB98}
signals that the bonds in the basal triangle of the molecule
might be rather soft. The other four susceptibility
measurements, \figref{F-9}, which are obtained from $B=3.5$~T to
$5$~T, can be better approximated by a second set of parameters:
$J/k_B=-3.2$~K, $J'/k_B=-3.1$~K, $j/k_B=1.5$~K, $j'/k_B=0$, and
$D/k_B=-3.2$~K (parameterization \textbf{2}).  These parameters
differ from those for $B=0.5$~T in two ways: (1) the biquadratic
contribution is 7~\% smaller, (2) the anisotropy coefficient is
three times larger.

\subsection{High-field properties of \nimo}

Although the above parameterization describes the low-field data
up to $B\approx 5$~T reasonably well, it fails to reproduce the
magnetization data at higher fields. The dramatic increase of
the field spacings between adjacent magnetization steps at 4.5,
8.9, 20.1, and 32~T cannot be explained by small changes of the
anisotropy $D$ or the coupling $J$. The third field spacing is
about twice the first one, which using relation \fmref{E-1-2}
suggests an exchange coupling that is about 1.25 of the original
one. In the following we therefore investigate a model where
only the exchange parameters are allowed to depend on field.
This is of course a simplification since all parameters of the
Hamiltonian ($J$, $j$, $D$, and $g$) should be modified in a
varying field. For our purpose we assume a hypothetical
exponential dependence of $J$ on the absolute value of the
external magnetic field
\begin{eqnarray}
\label{E-3-3-1}
J(B)
&=&
J_0
\exp\left(
\frac{|B|}{\gamma}
\right)
\ ,
\end{eqnarray}
which we motivate by a possible change of the overlap of those
orbitals that are involved in the superexchange. Such a change
could be caused by variations of the bond distances and angles;
the latter are known to have dramatic effects on the exchange
parameters. Based on the magneto-structural correlations for
Ni-O cubanes in Ref.~\onlinecite{HSH:IC95}, a change of the
exchange parameter in \nimo\ by the necessary amount would
correspond to a change of the bond angles by approximately one
quarter of a degree if the effect was attributed solely to bond
angle modifications. A literature survey of chemically-similar
Ni(II) cluster
compounds\cite{HSH:IC95,WHM:PRB98,EFK:P99,CMP:CC01,ABB:CEJ02,KWM:PRB03,YWH:P03,MHR:EJIC04}
reveals that $J$ and $D$ vary from at least -10 to
17.5~cm$^{-1}$ (-7 to 12.2~K) and -5.5 to 0.5 cm$^{-1}$ (-3.8 to
8.5~K), respectively. In these systems, bond angle is the
governing structural parameter due to the intimate relationship
between angle and magnetic orbital overlap.\footnote{In
  contrast, the Ni..Ni distances have only a small effect on the
  observed values of $J$.} There is a linear correlation between
the value of $J$ and Ni-O-Ni and Ni-O-O-Ni bond angles in
several nickel clusters. Although the transition metal centers
in these nickel clusters have different local environments, the
aforementioned range of $J$ and $D$ represents the variation in
physical parameters that can be accessed by chemical tuning in
these compounds. Magnetic field-induced effects are anticipated
to be in this range as well.

\begin{figure}[!ht]
\begin{center}
\epsfig{file=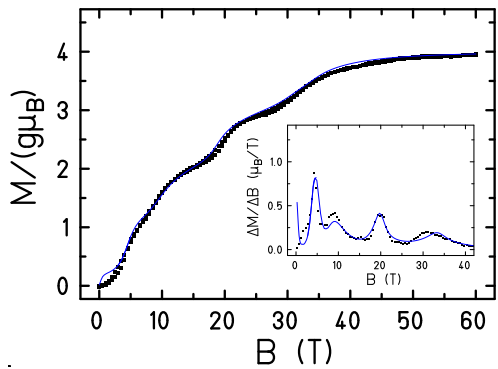,width=65mm} \vspace*{1mm}
  \caption[]{(color online) Magnetization of sample 2:
    Experimental data (NHMFL) are given by squares. The solid
    curve provides the best fit using a Hamiltonian with an
    exponentially field-dependent coupling. The simulated data
    are averaged over 100 orientations. The inset shows the
    experimental differential magnetization $\dint {\mathcal
      M}/\dint B$ as data points as well as the theoretical
    $\dint {\mathcal M}/\dint B$ (solid curve) using the same
    parameters as above. }
\label{F-10}
\end{center}
\end{figure}

The best fit using two exchange parameters was obtained for
$J/k_B=-4.2~\text{K}*\exp(B/96~\text{T})$,
$J'/k_B=-3.2~\text{K}*\exp(B/52~\text{T})$, $j/k_B=0.16$~K,
$j'/k_B=0.39$~K, $D/k_B=-8.9$~K, and $g=2.195$ together with the
paramagnetic impurity contribution already determined for sample
2 (parameterization \textbf{3}). This fit is shown in
\figref{F-10}. Except for small fields, where the two
aforementioned parameter sets are appropriate, this model
provides a rather good description of the magnetization over a
large field range. The assumption of an exponential dependence
is not essential; a polynomial dependence yields similar
results.

\subsection{EPR revisited}

The level scheme and possible EPR transitions can be shown more
closely using the parameters of the general Hamiltonian
\fmref{E-3-1-1}.

\begin{figure}[!ht]
\begin{center}
\epsfig{file=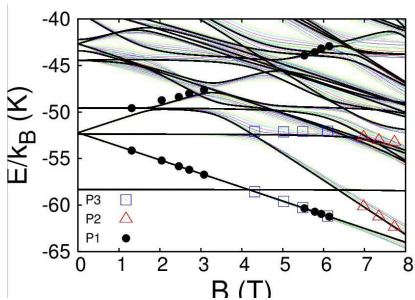,width=55mm,angle=0}
\vspace*{1mm}
\caption[]{(color online) The figure shows the Zeeman level
  splittings for the realistic Heisenberg model \fmref{E-3-1-1}
  (parametrization \textbf{2}) for several orientations of the
  field relative to the molecules. The various curves, which
  differ especially at avoided level crossings, represent
  relative angles of $10^\circ$.}
\label{F-11}
\end{center}
\end{figure}

In \figref{F-11} the Zeeman split levels are presented for
fields up to 8~T using parametrization \textbf{2}. The thick
curves show levels for an orientation of the field axis pointing
from the mid point of the basal triangle through the top Ni
center (Ni ion 1 in \figref{F-2}). The thinner curves show the
levels for orientations with relative angle steps of $10^\circ$
along a great circle through one of the basal Ni centers. The
variation of the level positions with orientation for a fixed
absolute value of the external field is especially large at
avoided level crossings.  Although the final EPR line is given
by averaging over all orientations one can already deduce from
\figref{F-11} that the EPR measurement is in full agreement with
the microscopic Hamiltonian. The attached symbols for the three
observed transitions match the calculated level spacings very
well.

\begin{figure}[!ht]
\begin{center}
\epsfig{file=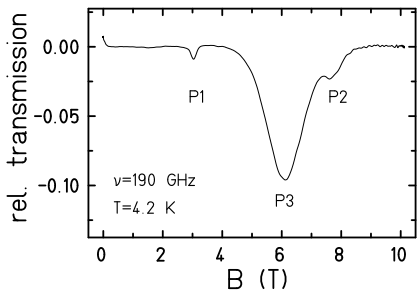,width=55mm,angle=0}
\vspace*{1mm}
\caption[]{Relative EPR transmission observed at $\nu=190$~GHz,
  which corresponds to an energy level difference of 9.12~K.:
  The dominant (allowed) transition P3 is rather broad, whereas
  the the weaker (forbidden) transition P1 results in a much
  sharper peak. The other weak (forbidden) transition P2, which
  is masked by P3, is broader than P1.}
\label{F-12}
\end{center}
\end{figure}

To some extent one can even explain the widths of the three
transitions, which are shown in \figref{F-12} for a frequency of
$\nu=190$~GHz and a temperature of $T=4.2$~K.  The transition
P1, which occurs between energy levels marked by filled circles
in \figref{F-11}, is rather narrow; this is connected to the
fact that variations of the field direction do not lead to
strong variations of the respective levels.  This also implies
that the mixing of $\op{S}_z$ eigenstates is weak and thus the
amplitude of this forbidden transition is small. P3 on the
contrary is an allowed transition which dominates the spectrum.
In the investigated field range it is given mainly by the
transition between levels marked by open squares in
\figref{F-11}, but transitions between other (unmarked) levels
also contribute. he rather broad line can in part be explained
by the rather strong variation of the upper marked level with
variations of the field direction. The transition P2 between
levels marked by open triangles in \figref{F-11} is stronger
than P1 but also broader, which can again be understood by
looking at the variation of levels.

\section{Conclusions}
\label{sec-5}

We have presented a comprehensive experimental and theoretical
investigation of the magnetic molecule \nimo. We find that the
main model parameters of the Hamiltonian (i.e. exchange and
anisotropy parameters) have a strong dependence on magnetic
field, an effect that may be accompanied by molecular
magnetostriction. All of our efforts to avoid this conclusion
lead to a blatant contradiction between theory and experiment.
The discovery of a dependence of Hamiltonian parameters on field
in \nimo\ is quite novel. Nevertheless, it may be a general
characteristic of magnetic cluster-based materials with strong
lattice coupling. It is well known that most materials
(including magnetic molecules) are not
rigid.\cite{SMW:PRB02,SDG:CEJ02,SDM:CEJ04,WCM:PRB,rai05} In
addition, only relatively few high-field magnetization studies
have been performed to date. Thus, it is of interest to discover
whether other magnetic molecules might display a similarly
strong dependence of the model parameters on magnetic field in
the regime above 20~T.

\begin{acknowledgments}
  This work was supported by the Deutsche Forschungsgemeinschaft
  (Grant No.  SCHN~615/8-1).  Ames Laboratory is operated for
  the U.S. Department of Energy by Iowa State University under
  Contract No.  W-7405-Eng-82. H.~N.  acknowledges the support
  by Grant in Aid for Scientific Research on Priority Areas (No.
  13130204) from MEXT, Japan and by Shimazu Science Foundation.
  Work at the University of Tennessee was supported by the
  Materials Science Division, Basic Energy Sciences, U.S.
  Department of Energy (DE-FG02-01ER45885) and the Petroleum
  Research Fund, administered by the American Chemical Society
  (38164-AC5).  The National High Magnetic Field Laboratory is
  supported by NSF Cooperative Agreement DMR-0084173 and by the
  State of Florida.  We thank K.~B\"arwinkel, S.~Hill,
  V.~Kataev, G.~Khaliullin, M.~Pederson, A.~Postnikov,
  R.~Saalfrank, J.~van Slageren, O.~Waldmann, and R.~Winpenny
  for helpful discussions.
\end{acknowledgments}


\end{document}